\documentclass[runningheads]{llncs}
\usepackage[T1]{fontenc}
\usepackage{amsmath}
\usepackage{amssymb}
\usepackage{nicefrac}
\usepackage{subcaption}
\usepackage{graphicx}
\usepackage{gensymb}
\usepackage{tabularx}
\usepackage{float}
\usepackage{placeins}
\graphicspath{{./images/}}
\begin{document}
\title{Global Context Is All You Need for\\ Parallel Efficient Tractography Parcellation}
%\title{Tractography parcellation gone wild:\\ Classifying everything, everywhere, all at once}
%

%\titlerunning{Tractography parcellation gone wild}
\titlerunning{Parallel Efficient Tractography Parcellation}
% If the paper title is too long for the running head, you can set
% an abbreviated paper title here
%
%\author{Anonymized Authors}  %% Added for anonymized MICCAI 2025 submission
%\authorrunning{Anonymized Author et al.}
%\institute{Anonymized Affiliations \\
%    \email{email@anonymized.com}}
\author{Valentin von Bornhaupt\inst{1}\and%\orcidID{0009-0005-2371-1459} \and
Johannes Grün\inst{1,2}\and%\orcidID{0000-0002-9154-3929} \and
Justus Bisten\inst{1,3}\and%\orcidID{0009-0007-8024-0803} \and
Tobias Bauer\inst{3,4}\and%\orcidID{0000-0002-0555-6214} \and
Theodor Rüber\inst{1,3,4}\and%\orcidID{0000-0002-6180-7671} \and
Thomas Schultz\inst{1,5}%\orcidID{0000-0002-1200-7248}
}
\authorrunning{V. von Bornhaupt et al.}
% First names are abbreviated in the running head.
% If there are more than two authors, 'et al.' is used.
%
\institute{University of Bonn, Bonn, Germany \and
  Center for X-ray and Nano Science CXNS, DESY, Hamburg, Germany \and
  University Hospital Bonn, Bonn, Germany \and
  German Center for Neurodegenerative Diseases (DZNE), Bonn, Germany \and
  Lamarr Institute for Machine Learning and Artificial Intelligence, Bonn, Germany\\
  \email{schultz@cs.uni-bonn.de}
}

\maketitle              % typeset the header of the contribution
\begin{abstract}
Whole-brain tractography in diffusion MRI is often followed by a parcellation in which each streamline is classified as belonging to a specific white matter bundle, or discarded as a false positive. Efficient parcellation is important both in large-scale studies, which have to process huge amounts of data, and in the clinic, where computational resources are often limited.
TractCloud is a state-of-the-art approach that aims to maximize accuracy with a local-global representation. We demonstrate that the local context does not contribute to the accuracy of that approach, and is even detrimental when dealing with pathological cases. Based on this observation, we propose PETParc, a new method for Parallel Efficient Tractography Parcellation. PETParc is a transformer-based architecture in which the whole-brain tractogram is randomly partitioned into sub-tractograms whose streamlines are classified in parallel, while serving as global context for each other. This leads to a speedup of up to two orders of magnitude relative to TractCloud, and permits inference even on clinical workstations without a GPU.
PETParc accounts for the lack of streamline orientation either via a novel flip-invariant embedding, or by simply using flips as part of data augmentation. Despite the speedup, results are often even better than those of prior methods. The  code and pretrained model will be made public upon acceptance.


\end{abstract}
\section{Introduction}
Whole-brain tractography is widely used in neurosurgical planning and scientific studies \cite{tobischCompressedSensingDiffusion2018,yangDiffusionMRITractography2021a,wilkeClinicalApplicationAdvanced2018}. It reconstructs a large number of streamlines from diffusion MRI data, with the goal of obtaining a comprehensive representation of the brain's white matter tracts. However, many false positive streamlines need to be removed from raw tractograms \cite{ismrm}, and the remaining streamlines need to be assigned to the corresponding anatomical tract for downstream analysis. This process is called tractogram parcellation, and significant research efforts have been focused on automating it. Early automated tools were either connectivity-based, where regions are defined that a streamline must intersect to be classified as a specific bundle \cite{wassermannWhiteMatterQuery2016}, or streamline-based, where a model is constructed for each bundle, and streamlines are classified based on their similarity to it \cite{garyfallidisRecognitionWhiteMatter2018}. All these methods require registration of the tractogram into a common space.

Recently, deep learning-based methods for tractogram parcellation have been proposed. Several issues arise when processing streamlines with neural networks, including their lack of orientation, implying that representations in forward or reversed order should be treated equivalently. Several streamline representations have been explored, including a mapping to 2D images, learned embeddings, computed features, and point clouds \cite{zhangDeepWhiteMatter2020,chenDeepFiberClustering2021,ngattailamTRAFICFiberTract2018,kumaralingamSegmentationWholeBrainTractography2022}. While these methods are effective, they still rely on registering the streamlines to a common space. Since registration is time-consuming and can be unreliable in pathological cases, TractCloud was recently proposed as a registration-free alternative \cite{xueTractCloudRegistrationfreeTractography2023}. TractCloud classifies each streamline separately, based on a local-global point cloud representation that is constructed individually for each streamline, a process which dominates the computational cost for inference.

We demonstrate with an ablation study that the local part of this representation does not contribute to TractCloud's accuracy. We believe that selecting local streamlines based on similarity is not particularly informative since it leads to high redundancy, and that it can even hurt generalization by introducing a dependency on the local streamline density. At the same time, omitting the need to individually construct a local context makes it possible to classify all streamlines in parallel, while they serve as a global context for each other.

This is the main idea behind PETParc, our novel transformer-based approach to Parallel Efficient Tractography Parcellation. Treating each streamline as a single token, self-attention determines which other streamlines are relevant for its classification. To preserve efficiency despite the quadratic time complexity of computing attention, we randomly partition the overall set of streamlines into subsets that are processed jointly, and that we refer to as sub-tractograms.

PETParc can account for the lack of orientation either via a novel, flip-invariant streamline embedding, or with a data-driven mapping that is learned with random flips as part of data augmentation. Both approaches achieve a speedup of up to two orders of magnitude relative to TractCloud, while results are similar, or even slightly better.
Even though it was trained on healthy adults, PETParc generalizes to different ages, imaging protocols, and health conditions. It outperforms TractCloud on challenging cases of individuals post-hemispherotomy, possibly due to normalization and heavy data augmentation. 
%%% Local Variables:
%%% mode: latex
%%% TeX-master: "main"
%%% End:

\section{Methods and Data}

\begin{figure}[t]
	\includegraphics[width=\linewidth]{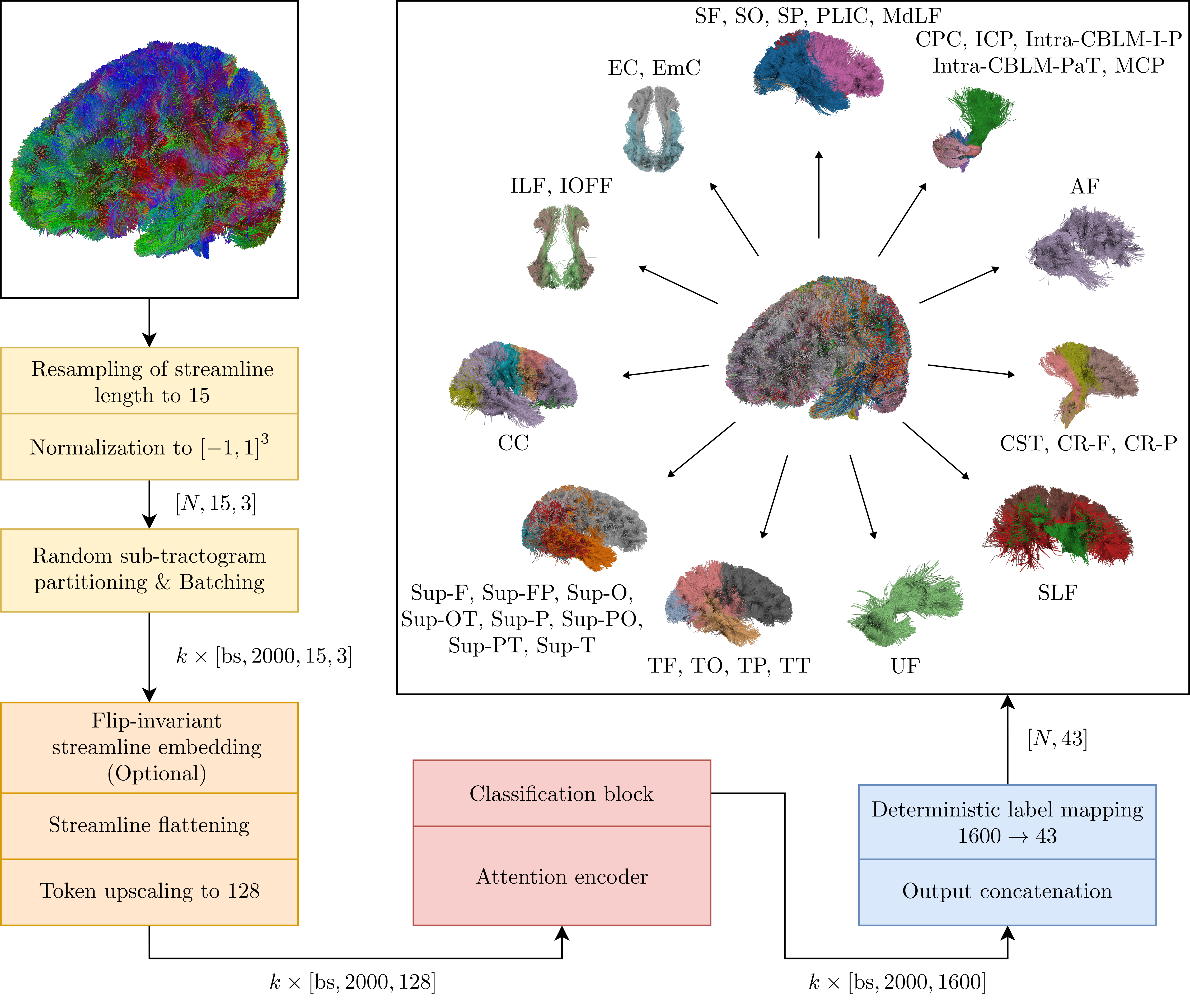}
	\caption{Our pipeline consists of the four stages preprocessing (yellow), embedding/token preparation (orange), transformer encoder/classifier (red), and postprocessing (blue). Tract names can be found in the paper by Zhang et al. \cite{AnatomicallyCuratedFiber}.}
	\label{fig:pipeline}
\end{figure}

To test our hypothesis that global context is all you need for registration-free tractography parcellation, we modified TractCloud so that it does not use any local context, resulting in the model variant TC\textsubscript{glo}. Except for the removal of local neighborhood features, the model architecture and hyperparameters remained the same as reported by Xue et al.\ \cite{xueTractCloudRegistrationfreeTractography2023}. This leads to encouraging results, motivating our novel method PETParc, which is illustrated in Figure~\ref{fig:pipeline}, and described in the remainder of this Section.

\subsection{Data} 
For training and quantitative evaluation, we use the dataset and labels provided by TractCloud \cite{xueTractCloudRegistrationfreeTractography2023}. This dataset defines 1600 clusters which are mapped to 42 bundles and one outlier class for evaluation. For external validation, we directly compare results from our method to TractCloud's on tractograms from four independently acquired subjects from the developing Human Connectome Project (dHCP) \cite{edwardsDevelopingHumanConnectome2022}, the Adolescent Brain Cognitive Development (ABCD) study \cite{volkowConceptionABCDStudy2018}, the HCP \cite{vanessenHumanConnectomeProject2012}, and the Parkinson's Progression Markers Initiative (PPMI) \cite{marekParkinsonsProgressionMarkers2018}.

We further evaluate the model on two million streamlines each that were created using a low-rank UKF \cite{gruenSpatiallyRegularizedLowrank2023a} for three individuals post-hemispherotomy, a procedure in which both hemispheres are surgically separated following severe damage in a single hemisphere. This study was approved by the Institutional Review Board of the medical faculty of the University of Bonn. Informed consent was obtained from all participants and/or their legal guardians.

\subsection{Preprocessing}
Streamlines are resampled to a fixed length of 15 points. Our architecture could support more detailed streamline representations, with the goal of better preserving high curvature, simply by adapting the initial tokenization. Since high accuracy is already achieved with the current, widely used choice \cite{xueTractCloudRegistrationfreeTractography2023,Zhang2024}, this is left as a potential topic of future investigation.

Coordinates are normalized to \([-1, 1]^3\) using affine min-max normalization along each axis. This ensures that all streamlines are transformed into a common space and scale, making the representation invariant to translation and scaling and eliminating the need for axis-specific scaling augmentations. The preprocessed streamlines are randomly partitioned into sub-tractograms of 2000 streamlines each, defined as the \texttt{context\_size}. This size needs to be sufficiently large that each randomly sampled sub-tractogram preserves suitable global context from the overall brain. However, it should not be much larger than necessary, due to the quadratic dependency of the transformer's computational effort on the \texttt{context\_size}. Multiple sub-tractograms are combined into batches. During evaluation, PETParc can process 512 sub-tractograms simultaneously, containing more than one million streamlines, using less than 10~GB of GPU memory.
\subsection{Flip-Invariant Embedding}
We introduce a novel, lossless streamline embedding that is flip-invariant in the sense that \( \text{emb} \left( v \right) = \text{emb} \left( \bar{v} \right) \), where emb denotes the embedding, $v\in \mathbb{R}^{15\times 3}$ a streamline and $\bar{v} = \left( v_{15}, \dots , v_1 \right)$ the same streamline in reversed order. The flip-invariant embedding is constructed as follows
\begin{align}
	\text{emb}_1 & := 0.5 \left(v_1+v_{15},\, v_2 + v_{14}, \dots, v_8 + v_8
	\right) \in \mathbb{R}^{8\times 3} \notag\\
\text{emb}_2 & := 0.5 \left(\vert v_1-v_{15}\vert, \, \vert v_2 - v_{14} \vert,\, \dots, \vert v_7 - v_9 \vert \right) \in \mathbb{R}^{7\times 3}\\
\text{emb}_3 & := (v_1 v_{14} + v_2 v_{15}, \, v_2 v_{13} + v_3 v_{14}, \dots, v_7
v_8 + v_9 v_8) \in \mathbb{R}^{7\times 3}, \notag
\end{align}
where addition, subtraction, and multiplication are performed component-wise. The embedding is now defined as \( \text{embedding}(v): = (\text{emb}_1, \text{emb}_2, \text{emb}_3) \in \mathbb{R}^{22\times 3}\). 
The components $\text{emb}_1, \text{emb}_2$ are based on the following identity: 
\begin{align}
	\{x,y\} = \left\{\frac{x+y \pm \vert x-y\vert}{2}\right\}.
\end{align}
We demonstrate that this embedding is lossless by observing that a streamline could be reconstructed by fixing either $v_1$ or $v_{15}$ as the first element, then using $\text{emb}_3$ to construct a linear system of equations. Due to flip invariance, this reconstruction is unique up to the original orientation. In some degenerate cases, such as subsequences $(v_i, v_{i+1}, \dots, v_{n-i}, v_{n-i+1})$ that satisfy $v_i = v_{n-i+1}$ or $v_{i+1} = v_{n-i}$, the system may lack a unique solution. However, these cases are unlikely to occur in practice. To achieve comparable numerical ranges, we apply $f(x) = \text{sign}(x)\sqrt{\vert x \vert}/2$ on all components of $\text{emb}_3$. As an alternative to the flip-invariant embedding, we apply flip augmentation during training, independently flipping each streamline with 0.5 probability. 

\subsection{Model Training}
The backbone of the model is a transformer encoder consisting of eight layers. Each layer uses single-head self-attention and a feed-forward network with 256 hidden units. We used a token dimension of 128, for which we observed optimal model accuracy. A dropout rate of $0.1$ is applied within each layer, along with layer normalization. The output from the transformer encoder is passed through a 256 hidden units classification block and a ReLU activation, followed by a second linear layer that outputs 1600 units. The model was trained using the Adam optimizer with a learning rate of $8.5 \cdot 10^{-4}$, a weight decay of $10^{-3}$ and a cosine annealing learning rate scheduler. Training was performed with a batch size of 64 for 50,000 epochs, employing a cross-entropy loss function. Training took approximately 20 hours for the flip-inv model and 19 hours for the flip-aug model. The 1600 clusters were used as labels during training, which was performed on a single NVIDIA A40 GPU with 48GB of memory.
Each sample in a batch consisted of a random sub-tractogram of \texttt{context\_size = 2000} streamlines from a single training subject. Following TractCloud, rotation angles were applied uniformly from $[-45\degree, 45\degree]$ for the left-right axis, and $[-10\degree, 10\degree]$ for both the anterior-posterior and superior-inferior axes. Since our approach does not require recalculating local or global features after augmentation, these rotations were applied in every epoch for every sub-tractogram, resulting in approximately $3.2$ million rotation augmentations. Gaussian noise with $\sigma=0.001$ was added to each point of the streamlines. The streamlines were then re-normalized to $[-1, 1]^3$ to ensure that the scale of the transformer input remained consistent despite the effects of rotation and noise. Due to the inherent permutation invariance of the transformer architecture \cite{vaswani2023attentionneed}, the order of streamlines within a sub-tractogram does not affect their classification. Moreover, the architecture supports evaluation with arbitrary context sizes.
%%% Local Variables:
%%% mode: latex
%%% TeX-master: "main"
%%% End:

\section{Experiments and Results}
\subsection{Performance on test dataset}
\begin{table}[tb]
  \begin{minipage}{\textwidth}
    \caption{Comparison of the flip-augmented and flip-invariant version of PETParc with TC\textsubscript{\textit{glo+loc}} and TC\textsubscript{\textit{glo}} on both the test dataset and the synthetic transform augmented (STA) dataset. Asterisks indicate values taken from \cite{xueTractCloudRegistrationfreeTractography2023}. Subscript numbers indicate the \texttt{context\_size} used for the respective model.}
    \centering
    \begin{tabularx}{\textwidth}{|l|*{3}{>{\centering\arraybackslash}X|}*{1}{>{\centering\arraybackslash}X|}}
    \hline
    & \multicolumn{4}{c|}{Test-Data -- 200,000 Streamlines} \\
    \cline{2-5}
    & Acc [\%] & F1 [\%] & Time [s] & CPU-Only [s] \\
    \hline
    DeepWMA & 90.29 * & 88.12 * & n.a. & n.a. \\
    DCNN++ & 91.26 * & 89.14 * & n.a. & n.a. \\
    TC\textsubscript{\textit{glo+loc}} & 91.99 * & 90.10 * & 33.64 & 310.35 \\
    TC\textsubscript{\textit{glo}} & 92.13 & 90.27 & 12.42 & 253.64 \\
    PETParc Flip-Inv & 93.90 & 92.39 & 0.28 & 6.88 \\
    PETParc Flip-Aug & \textbf{94.75} & \textbf{93.46} & \textbf{0.26} & \textbf{6.84} \\
    \hline
    \end{tabularx}
    \label{tab:acc}
  \end{minipage}

  \vspace{0.8em} 

  \begin{minipage}{\textwidth}
    \centering
    \begin{tabularx}{\textwidth}{|l|*{3}{>{\centering\arraybackslash}X|}*{1}{>{\centering\arraybackslash}X|}}
    \hline
    & \multicolumn{4}{c|}{STA-Data -- 6.2 Mio Streamlines} \\
    \cline{2-5}
    & Acc [\%] & F1 [\%] & Time [s] & CPU-Only [s] \\
    \hline
    TC\textsubscript{\textit{glo+loc}} & 91.69 * & 89.65 * & 1027.80 & 8869.82 \\
    TC\textsubscript{\textit{glo}} & 91.71 & 89.66 & 352.13 & 8005.17 \\
    PETParc Flip-Inv & 93.30 & 91.55 & 5.48 & 235.88 \\
    PETParc Flip-Aug\textsubscript{500} & 93.35 & 91.62 & \textbf{3.16} & \textbf{130.82} \\
    PETParc Flip-Aug\textsubscript{2000} & \textbf{94.26} & \textbf{92.80} & 5.41 & 232.99 \\
    \hline
    \end{tabularx}
  \end{minipage}
\end{table}
An initial insight from comparing TC\textsubscript{\textit{glo+loc}}, a state-of-the-art registration-free tractography parcellation method \cite{xueTractCloudRegistrationfreeTractography2023} to its variant TC\textsubscript{\textit{glo}} (Table~\ref{tab:acc}) is that omitting the local context saves considerable time without hurting accuracy. Results of the registration-based alternatives DeepWMA \cite{zhangDeepWhiteMatter2020} and DCNN++ \cite{Xu2019} illustrate that omitting registration is not just a matter of saving time, but also affords higher accuracy. We could not directly compare timings ourselves, but prior work reported that TractCloud runs faster than DeepWMA and DCNN++ even when disregarding the time required for registration \cite{xueTractCloudRegistrationfreeTractography2023}.

Table~\ref{tab:acc} also shows that both variants of PETParc yield even better results, and are much faster. The flip-augmented model leads to slightly better F1 and accuracy scores than the flip-invariant model. Moreover, the flip-augmented approach requires only 5.41~s to classify 6.2 million streamlines on a single NVIDIA A40 GPU, demonstrating a remarkable improvement in efficiency. In contrast, TC\textsubscript{\textit{glo+loc}} takes 1027.80~s to parcellate the same tractogram, requiring 190 times more time. Reducing PETParc's context size to 500, the same as TractCloud, further reduces computation times at a minor loss in accuracy.
During evaluation, PETParc utilized a maximum of 9.42~GB of GPU memory for context size 2000, 3.54~GB for context size 500. To ensure a fair comparison, we tested different batch sizes for TractCloud. Even on a 6th-gen i7 CPU without a GPU, all PETParc variants processed 6.2M streamlines in under four minutes.

\subsection{Robustness against streamline flipping and random sub-tractogram partitioning}
\begin{table}[tb]
  \caption{Evaluation of the stability of flip-invariant and flip-augmented models against random flips, random sub-tractogram partitioning (RSP) and their combination over $n=20$ experiments.}
  \centering
  \begin{tabularx}{\textwidth}{|l|*{2}{>{\centering\arraybackslash}X|}*{2}{>{\centering\arraybackslash}X|}}
  \hline
  & \multicolumn{2}{c|}{Flip-Invariant} & \multicolumn{2}{c|}{Flip-Augmented} \\
  \cline{2-5}
  & Mean Acc & Std. Acc & Mean Acc & Std. Acc \\
  \hline
  Without  & 93.96 & $0.0$ & 95.14 & $0.0$ \\ 
  Random Flips & 93.96 & $0.0$ & 95.05 & $6.9 \times 10^{-4}$ \\
  RSP & 93.90 & $8.0 \times 10^{-4}$ & 94.96 & $9.7 \times 10^{-4}$ \\
  RSP + Random Flips & 93.90 & $8.0 \times 10^{-4}$ & 94.97 & $8.6 \times 10^{-4}$ \\
  \hline
  \end{tabularx}
  \label{tab:flip-inv-aug}
\end{table}
We evaluate the robustness of our invariant and augmented models to random streamline flipping and partitioning, using $n=20$ runs on the first TractCloud test subject (see Table~\ref{tab:flip-inv-aug}). As expected, the flip-invariant model remains unaffected by random flips, maintaining a constant accuracy of 93.96\%. Random sub-tractogram partitioning (RSP) slightly lowers its accuracy (93.90\%) with minimal variance. The flip-augmented model achieves higher accuracy (95.1\%) but is slightly affected by flips (95.05\%) and RSP (94.96\%), though with negligible variance. Overall, both models are stable, with the flip-augmented model achieving better accuracy but showing minor sensitivity to flips. For practical purposes, we find it sufficient to learn flip invariance through augmentations.

\subsection{Qualitative evaluation}
\begin{figure}
  \centering
  \includegraphics[width=\textwidth]{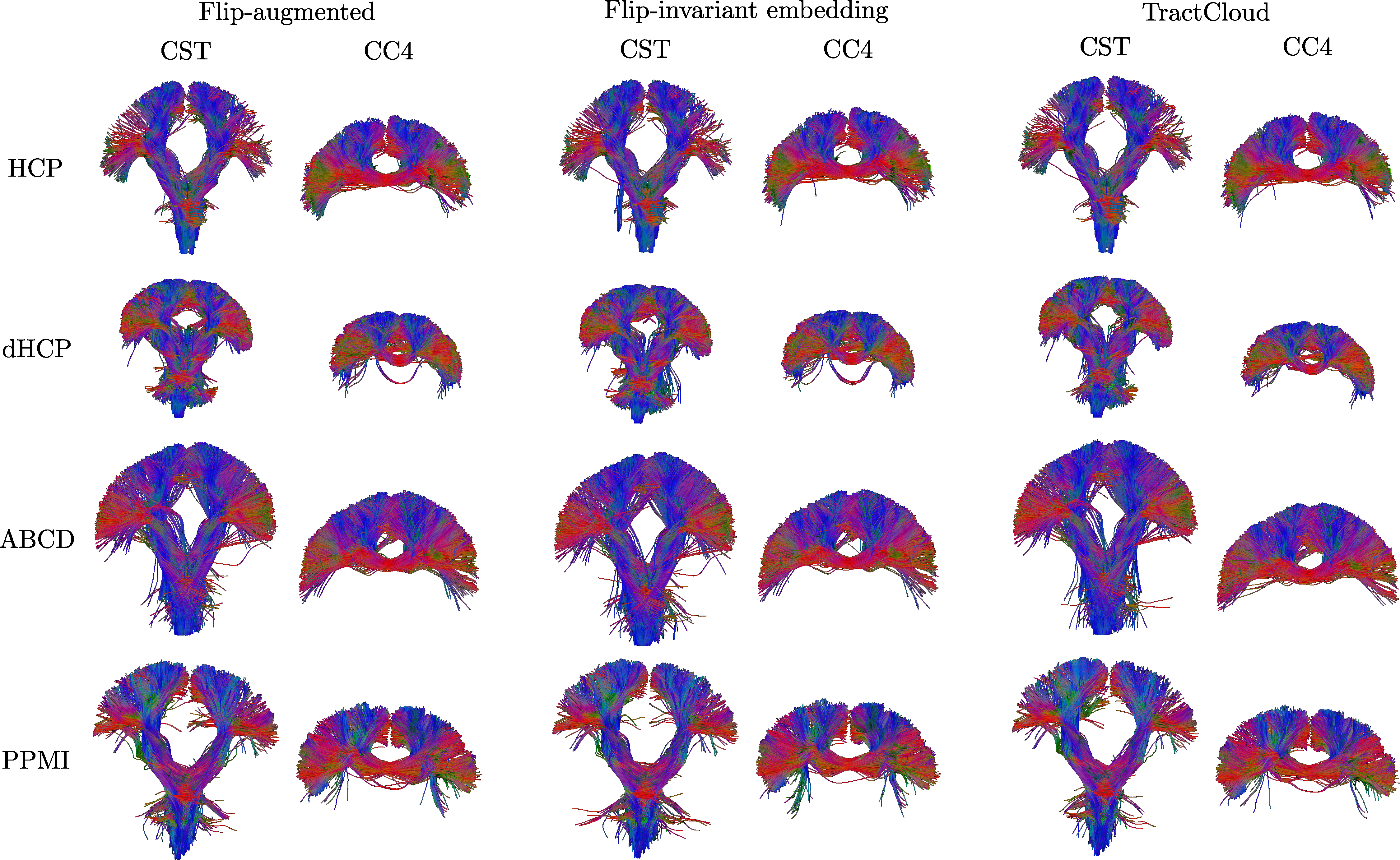}  % image.png should be inside the 'graphics' folder
  \caption{Comparison of the flip-augmented and flip-invariant versions of PETParc to TractCloud on four subjects from different studies. All models generalize well to different measurement schemes, different ages, and different health status.}
  \label{fig:opendatasets}
\end{figure}

To evaluate the generalization capability of the models, we tested them on subjects from the ABCD, PPMI, HCP, and dHCP datasets, comparing PETParc with TC\textsubscript{\textit{glo+loc}}. All models yield promising results, even for the challenging dHCP subject (see Figure \ref{fig:opendatasets}).

\begin{figure}
  \centering
  \includegraphics[width=\textwidth]{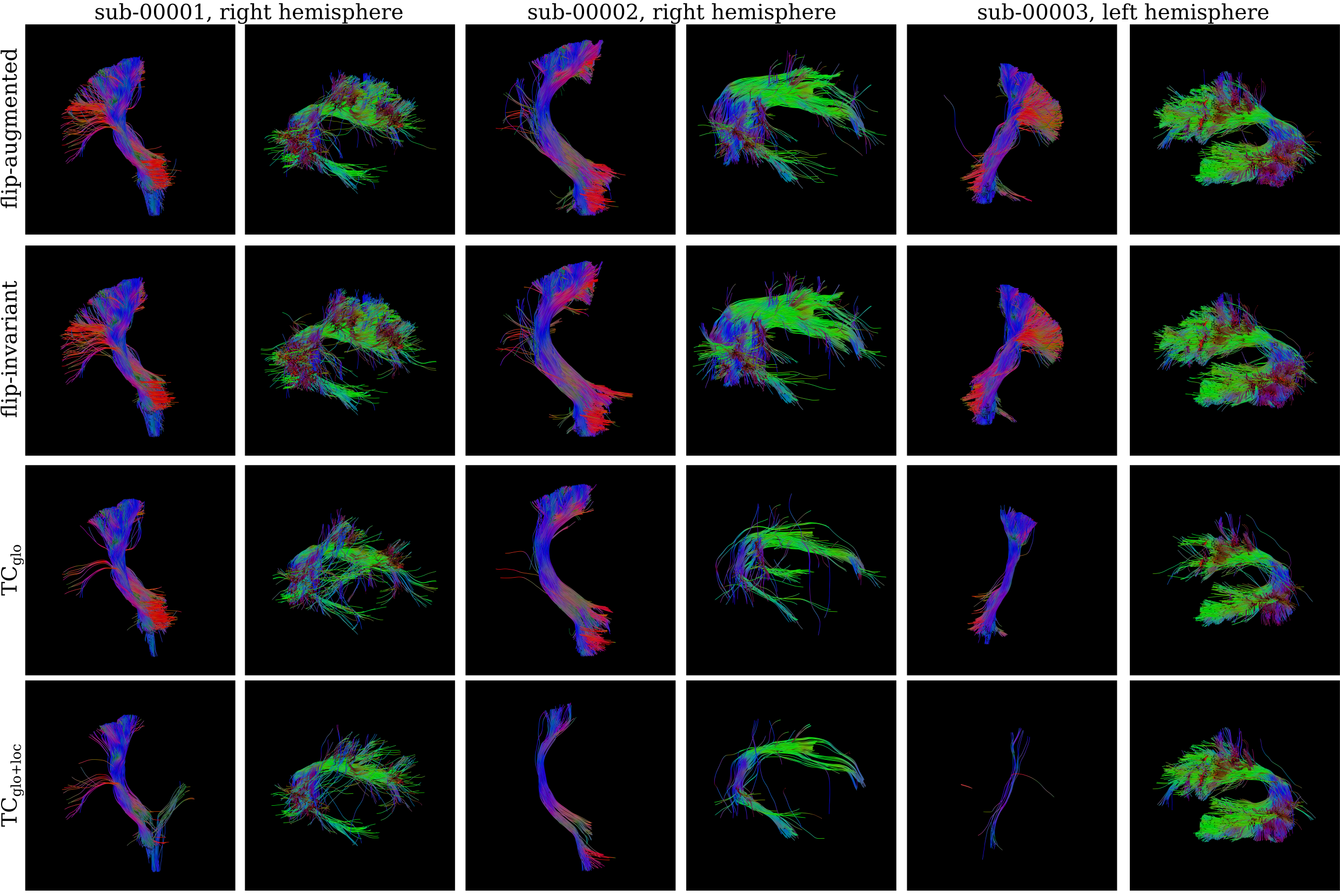}  % image.png should be inside the 'graphics' folder
  \caption{
    Parcellation results of the corticospinal tract (CST) and arcuate fasciculus (AF) in three individuals post-hemispherotomy. Both variants of PETParc yield more complete parcellations compared to TC\textsubscript{\textit{glo+loc}} and TC\textsubscript{\textit{glo}}.}
  \label{fig:hemi}
\end{figure}

To assess generalizability with respect to severe lesions, we analyzed the parcellated tractograms from three individuals post-hemispherotomy (see Figure~\ref{fig:hemi}). Registration-free models are highly useful in this case, as registration often fails for images containing severe lesions. PETParc parcellations of all the corticospinal tracts (CSTs) and the arcuate fasciculus (AF) in Subjects~1 and~2 appear to be more complete compared to TC\textsubscript{\textit{glo+loc}}. TC\textsubscript{\textit{glo}} yields slightly more complete parcellations compared to TC\textsubscript{\textit{glo+loc}}, in line with our intuition that relying on a local context can hurt generalization. Our more comprehensive augmentations might explain the remaining gap towards PETParc's results.

\section{Conclusion}
Building on TractCloud, a state-of-the-art method for tractography parcellation, we investigated the relevance of local context information. The fact that a simplified variant TC\textsubscript{\textit{glo}} achieves comparable accuracy, and even improved generalization to cases with severe pathology, suggests that including a local context for each individual streamline is not just computationally expensive, but in fact not useful for streamline classification.

Following this insight, we introduced PETParc, the first transformer based method for ultrafast, parallel tractogram parcellation, leveraging only global context. PETParc outperformed TC\textsubscript{\textit{glo+loc}} in both accuracy and macro F1 score while achieving up to two orders of magnitude faster inference. This speedup is enabled by our massively parallelized approach, which classifies all streamlines within a sub-tractogram at once. As part of PETParc's design, we explored a novel flip-invariant streamline embedding that formally guarantees that reversing a streamline does not affect its classification. However, comparing results to streamline flipping during data augmentation suggests that data-driven embeddings can implicitly learn flip-invariance to a large extent.

Finally, we demonstrated the robustness of our approach on individuals with severe lesions, rendering PETParc highly suitable for clinical applications, studies involving large-scale tractography, and studies of lesioned brains.

\FloatBarrier
%%% Local Variables:
%%% mode: latex
%%% TeX-master: "main"
%%% End:

\begin{credits}
  \subsubsection{\ackname} Funded by the German Federal Ministry of Education and Research (projects ``BNTrAinee'', funding code 16DHBK1022, and ``epi-center.ai''); by the Neuro-aCSis Bonn Neuroscience Clinician Scientist Program (2024-12-07); and by the TRA Life and Health (University of Bonn) as part of the Excellence Strategy of the federal and state governments. We gratefully acknowledge access to the Marvin cluster of the University of Bonn.

\subsubsection{\discintname}
The authors have no competing interests to declare that are
relevant to the content of this article.
\end{credits}

%\clearpage
\bibliographystyle{splncs04}
\bibliography{bibfile}

\end{document}